\documentclass[rnote]{aa} 

\usepackage{graphicx}
\usepackage{txfonts}
\begin{document}

   \title{Importance of the H$_2$ abundance in protoplanetary disk ices for the molecular layer chemical composition}
\titlerunning{H$_2$ in the molecular layer}


   \author{V. Wakelam\inst{1}
          \and
          M. Ruaud\inst{1}
          \and 
          F. Hersant\inst{1}
          \and
          A. Dutrey\inst{1}
          \and
          D. Semenov\inst{2}
          \and 
          L. Majumdar\inst{1,3}
          \and 
          S. Guilloteau\inst{1}
          }

   \institute{Laboratoire d'astrophysique de Bordeaux, Univ. Bordeaux, CNRS, B18N, all\'ee
Geoffroy Saint-Hilaire, 33615 Pessac, France
             \email{valentine.wakelam@u-bordeaux.fr}
        \and 
        Max Planck Institute f\"{u}r Astronomie, K\"{o}nigstuhl 17, 69117 Heidelberg, Germany
         \and
         Indian Centre for Space Physics, 43 Chalantika, Garia Station Road, Kolkata, 700084, India     }

   \date{Received xxxx; accepted xxxx}

 
  \abstract
   {Protoplanetary disks are the target of many chemical studies (both observational
and theoretical) as they contain the building material for planets. Their large vertical
and radial gradients in density and temperature make them challenging objects for
chemical models. In the outer part of these disks, the large densities and low
temperatures provide a particular environment where the binding of species onto the
dust grains can be very efficient and can affect the gas-phase chemical composition.
}
   {We attempt to quantify to what extent the vertical abundance profiles and the integrated
column densities of molecules predicted by a detailed gas-grain code are affected by the
treatment of the molecular hydrogen physisorption at the surface of the grains.}
   {We performed three different models using the Nautilus gas-grain code. One model
uses a H$_2$ binding energy on the surface of water (440 K) and produces strong sticking
of H$_2$. Another model uses a small binding energy of 23 K (as if there were already a
monolayer of H$_2$), and the sticking of H$_2$ is almost negligible. Finally, the remaining
model is an intermediate solution known as the encounter desorption mechanism.}
   {We show that the efficiency of molecular hydrogen binding (and thus its
abundance at the surface of the grains) can have a quantitative effect on the predicted
column densities in the gas phase of major species such as CO, CS, CN, and HCN.}
   {}


   \maketitle
%

\section{Introduction}

 Protoplanetary gas and dust disks encountered around low-mass TTauri stars are the progenitors of planetary systems.  The 2D physical structure of these objects has been the subject of many observational and theoretical studies, which agree on several key points. Disks exhibit important radial and vertical density and temperature gradients. Temperatures range from 10~K and below around the cold mid-plane to thousands of Kelvin in the very inner disk (less than radius $r=10$ au from the central star) in the upper atmosphere. The densities are small in the upper layers while they can reach $10^{12}$~cm$^{-3}$ and more in the mid-plane in the very central region. Moreover, photo-dissociation processes play an important role owing to the emission of the central star in the UV and X-ray wavelength domains \citep{2013ChRv..113.9016H,2014prpl.conf..317D}. 
As such, protoplanetary disks are very challenging sources for astrochemical models as chemistry and physics are intimately linked and chemical processes have to be properly taken into account to retrieve disk physical
properties. To study the impact of chemistry on observable molecular species, many models have been developed over the years with increasing sophistication \citep[see][and references therein]{2014prpl.conf..317D}.  

In the regions of the disks where the density is high and the temperature low, depletion of H$_2$ on interstellar grains becomes a problem for chemical models. The binding energy of H$_2$ on water ice is around 440~K \citep{2007ApJ...668..294C}. As we show, assuming such large binding energy and in the absence of a detailed microscopic simulation of the chemistry, the depletion of H$_2$ on the surface would remove a large fraction of the total mass of the gas. However, after the formation of the first monolayer of H$_2$, the sticking of H$_2$ over itself is rather inefficient.  \citet{2007ApJ...668..294C} proposed an H$_2$ onto H$_2$ binding energy of 23~K. Assuming only this value for the binding energy of H$_2$ underestimates the abundance of H$_2$ on the surface by orders of magnitude \citep{2015A&A...574A..24H}. Most - if not all - publications about disk chemical modeling do not describe their way of solving the problem. \citet{2015A&A...574A..24H} have proposed to introduce in gas-grain codes a new reaction: s-H$_2$ + s-H$_2$ $\rightarrow$ s-H$_2$ + H$_2$, which takes into account the facile desorption of H$_2$ when adsorbed on itself \citep{2007ApJ...668..294C}. This new mechanism is called "encounter desorption" by the authors and its efficiency has been tested against microscopic Monte Carlo simulations, which explicitly follow the landing of species on the surface. 

 In this paper, we show to what extent the computed species abundances are sensitive to the abundance of H$_2$ on the grains and so the treatment of its sticking on the grain surfaces.


\section{Model description}
To model the chemistry in protoplanetary disks, we have used the gas-grain code Nautilus \citep{2015A&A...579A..82R,2015MNRAS.447.4004R} with a 1D vertical structure at a given radius in the disk. The physical structure is described in section~\ref{physical_model}, while the chemical model is presented in section~\ref{chemical_model}.

\subsection{Disk physical model}\label{physical_model}

The disk structure is similar to that used in \citet{2015A&A...579A..82R} and \citet{2009A&A...493L..49H} (with a refinement for the temperature profile). We briefly summarize below  some key points. 

The vertical temperature profile, at a given radius $r$ from the central star, is computed using the formalism originally introduced by \citet{2003A&A...399..773D} and modified by \citet{2014ApJ...788...59W}. In this approximation, the temperature is assumed constant above four scale heights ($h$). 
Below four scale heights, the vertical temperature $\rm T(z)$ is computed as
\begin{equation}
	\rm T(z) = T_{mid} + (T_{atm} - T_{mid})  \Bigg[\sin{\Bigg(\frac{\pi z}{2z_q}\Bigg)}\Bigg]^{2\delta},
\end{equation}
where $\rm T_{mid}$ and $\rm T_{atm}$ are respectively the temperatures in the mid-plane and at 4$h$, $\rm z_q$ is the altitude at 4$h$ and $\delta$ is the steepness of the profile. 
The disk is in hydrostatic equilibrium and the vertical density distribution is computed with Eq.~1 of \citet{2015A&A...579A..82R}. The surface density is taken as 
\begin{equation}
\rm \Sigma(r) =  \Sigma_{r_o} \times (r/r_o)^{-1.5}.
\end{equation} 
Here $\rm \Sigma_{r_o}$ is linked to the total disk mass $\rm M_{disk}$ and the outer disk radius  $\rm r_{out}$ by the relation $\rm \Sigma_{r_o} =  \frac{M_{disk} r_o^{-1.5}}{4\pi \sqrt{r_{out}}}$.
In this simulation, the total mass of  the disk $\rm M_{disk}$ is 0.03 M$_{\odot}$ \citep{2011A&A...529A.105G}, which corresponds to $\rm \Sigma_{r_o} = 0.8$ g/cm$^2$ for $\rm r_o = 100$ au. 

\begin{figure}
\includegraphics[width=0.8\linewidth]{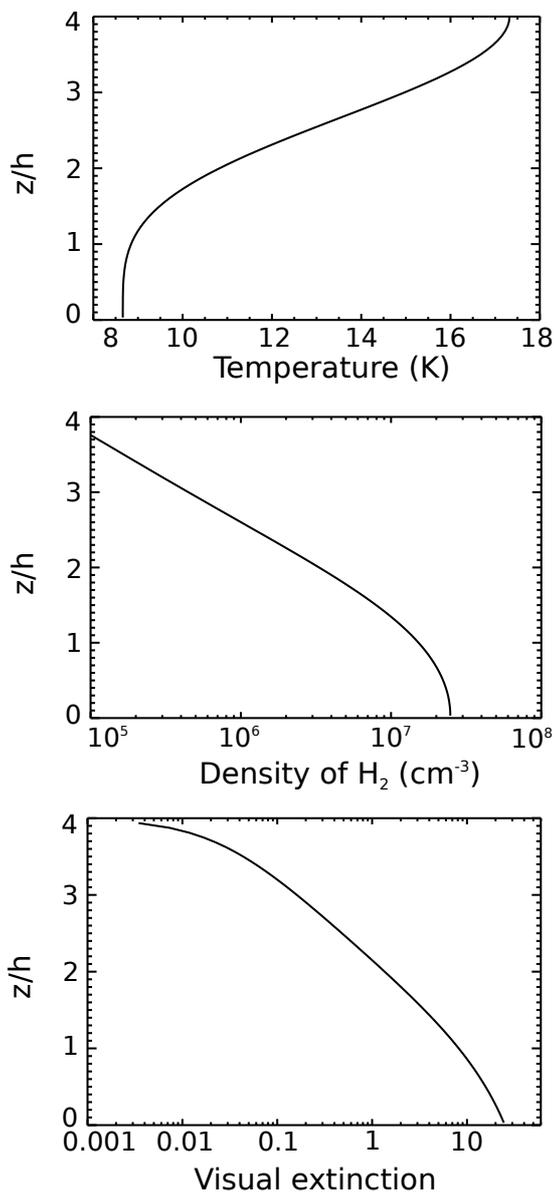}
\caption{Temperature, density (of H$_2$), and visual extinction as a function of z/$h$ at 300 au. \label{disk_verticalprofil}}
\end{figure}

The vertical visual extinction profile is computed from the hydrostatic density structure assuming a conversion factor of $\rm N_H / A_v = 1.6\times 10^{21}$ \citep{1989MNRAS.237.1019W}. The H$_2$, CO, and N$_2$ photo-dissociation rates are computed using self-shielding functions from \citet{1996A&A...311..690L}, \citet{2009A&A...503..323V}, and \citet{2013A&A...555A..14L}, respectively. The UV flux coming from the star is assumed to have the same spectrum as the UV interstellar field. Since we do not do a full 3D treatment of the radiative transfer, the UV flux at a radius r from the star is assumed to be absorbed at the surface of the disk and half of the flux diffuses in the vertical direction towards the mid-plane
\begin{equation}
\rm f_{UV} = \frac{f_{UV100au}/2}{\Big(\frac{r}{100 au}\Big)^2 + \Big(\frac{4h}{100  au}\Big)^2},
\end{equation}
where $\rm f_{UV100au}$ is the UV flux factor at 100 au as usually reported in observational papers.\\
For the purpose of this study, we have used physical parameters similar to the ones of the protoplanetary disk around the DMTau star: $\rm f_{UV100au}$ = 410 $\chi_0$\footnote{The UV flux is given in units of the \citet{1978ApJS...36..595D} interstellar UV field.} \citep{2004ApJ...614L.133B}, $\rm M_*$ = $0.53$ M$_{\odot}$ \citep{2000ApJ...545.1034S}, $\rm M_{disk}$ = 0.03 M$_{\odot}$ \citep{2011A&A...529A.105G}, and $\rm r_{out}$ = 800 au \citep{1998A&A...339..467G}. The dust-to-gas mass ratio is assumed to be $10^{-2}$. The atmospheric and mid-plane temperatures at 300 au are taken as 17.3 and 8.6 K respectively. The dust temperature is assumed to be equal to that of the gas. This assumption is very likely not true in the upper atmosphere of the disk \citep{2007prpl.conf..751B}. The conclusions of this paper, which concern mostly the regions of the disk where dust plays a role in the chemistry, are unaffected by this simplification. The atmospheric and mid-plane temperatures were obtained at 300 au assuming a mid-plane and an atmospheric temperature of 10 and 30 K, respectively, at 100 au and with a scaling $T_{100}(r/100 {\rm au})^{-0.5}$  \citep{2007A&A...467..163P}. 
The stiffness of the vertical temperature profile ($\delta$) is taken to be 2 but this choice is not crucial for the purpose of this study \citep[see also][]{2015A&A...579A..82R}. The parameters used to compute the physical structure are summarized in Table \ref{tab:disk}. The computed vertical structure at 300 au of the disk is shown in Fig.\ref{disk_verticalprofil} as a function of z/$h$. At this radius, the scale height $h$ is 41.5 au and $\rm f_{UV}$ is 17. 

\begin{table}
\caption{Physical structure of the disk and stellar parameters}
\begin{center}
\begin{tabular}{lc}
\hline
\hline
Parameter  & Value \\
\hline
Stellar Mass & M$_*$  = 0.53 M$_{\odot}$ \\
Stellar UV flux &  $\rm f_{UV100AU}$ = 410 $\chi_0$ \\
\hline 
Dust-to-gas Ratio & 10$^{-2}$ \\
Total Disk Mass & M$_{disk}$ = 0.03 M$_{\odot}$ \\
Surface density & $\rm \Sigma_{100}  = 0.8 $g/cm$^2$   \\
Outer Radius & $\rm R_{out}  = 800$ au   \\
Upper Surface Temperature (at 300 au) & $\rm T_{atm}  $= 17.3 K  \\
Mid-Plane Temperature (at 300 au) & $\rm T_{mid}$ = 8.6 K  \\
 $\rm z_q$ & 4h    \\
 $\rm \delta$ & 2 \\
\hline
\end{tabular}
\end{center}
\label{tab:disk}
\end{table}

\subsection{Chemical model}\label{chemical_model}

To compute the chemistry in the vertical direction at a specific radius of the disk, we have used the gas-grain code Nautilus described in \citet{2016MNRAS.459.3756R}. This code allows the computation of the chemical composition in the gas-phase and at the surface of interstellar grains. The gas phase processes are determined by the public chemical network kida.uva.2014 \citep{2015ApJS..217...20W}. While colliding with grains, the species in the gas phase can be physisorbed on the surface.  Species at the surface of the grains (with a unique size of 0.1$\mu$m) can diffuse and undergo chemical reactions following the "rate equation" method from \citet{1992ApJS...82..167H}. We use Nautilus in its  two-phase mode, which means that we make no distinction between the bulk of the mantle and its surface. The desorption mechanisms are 1) thermal desorption \citep{1992ApJS...82..167H}, 2) cosmic-ray induced desorption, 3) chemical desorption and 4) photo-desorption. The cosmic-ray induced desorption is based on the hypothesis that high-energy iron nuclei (from the cosmic-ray particles) colliding with a grain will heat it up to a certain temperature for a certain amount of time. Following \citet{1985A&A...144..147L} and \citet{1993MNRAS.263..589H}, we assume that the grains are entirely heated up to 70~K for approximately $10^{-5}$~s every $3\times 10^{13}$~s. The chemical desorption is produced by the extra energy released by exothermic reactions at the surface of the grains. We have followed the formalism by \citet{2007A&A...467.1103G} and assumed that only 1\% of the products of each reaction are evaporated through this mechanism. The surface network and the species binding energies are the same as in \citet{2016MNRAS.459.3756R}. The diffusion energy of all species is assumed to be 0.5 of their binding energy except for atomic hydrogen, whose diffusion energy is taken as 230~K \citep{2007MNRAS.382.1648A}. The value of the ratio between diffusion and desorption energies we have chosen is within the range of values used by \citet{2016A&A...585A.146M} to reproduce their experiments on O and N diffusion/desorption rates. This model takes into account the competition between diffusion, reaction, and evaporation for the computation of the probability of surface reaction. Photo-desorption by direct and indirect UV photons is also included with a constant yield for all species of $10^{-4}$ molecules per photon \citep[see][for details]{2016MNRAS.459.3756R}.\\

To compute the chemistry in the protoplanetary disk, we first calculate the chemical composition (gas and ices) assuming dense cloud conditions for $10^6$~yr and using the elemental abundances listed in \citet{2011A&A...530A..61H}. For the cloud, everything is initially in the atomic form (except H$_2$) with an oxygen elemental abundance of $2.4\times 10^{-4}$ compared to the total proton density. The gas and dust temperature is 10~K, the H$_2$ density is $10^4$~cm$^{-3}$, the visual extinction is 15 mag and the cosmic-ray ionization rate is $1.3\times 10^{-17}$~s$^{-1}$. The final cloud composition is assumed to be the initial chemical conditions of the disk. We note that as soon as the age of the initial cloud is assumed to be less than or equal to $10^6$~yr, the disk chemical composition does not depend on the initial cloud composition. Even starting with atoms will very rapidly produce the same abundances in the disk. Three models were run. Model 1 includes the encounter desorption mechanism proposed by \citet{2015A&A...574A..24H}. \citeauthor{2015A&A...574A..24H} have proposed introducing in grain chemical networks the reaction s-H$_2$ + s-H$_2$ $\rightarrow$ s-H$_2$ + H$_2$, which mimics the encounter of two H$_2$ molecules on the same surface site. The rate of this reaction (in cm$^{-3}$s$^{-1}$) is then computed by the formula
\begin{equation}
\rm R_{H_2H_2} = \frac{1}{2} k_{H_2H_2}n_s(H_2)n_s(H_2)\kappa(H_2),
\end{equation}
with $\rm k_{H_2H_2}$ (in cm$^3$s$^{-1}$) the sum of the diffusion rates of two H$_2$ into the same site as defined by \citet{1992ApJS...82..167H}, $\rm n_s(H_2)$ the density of H$_2$ on the surface (in cm$^{-3}$) and $\rm \kappa(H_2)$ the probability that H$_2$ desorbs rather than diffuses \citep[Eq. 2 of][]{2015A&A...574A..24H}. This probability compares the sum of all the desorption rates of H$_2$ (through all mechanisms) to the diffusion rate of H$_2$ on an H$_2$ substrate (i.e. all computed with a H$_2$ binding energy of 23~K). Even though the binding energy of H$_2$ on H$_2$ is uncertain, the diffusion rate of H$_2$ is usually much smaller than the desorption rates at 10~K so that this probability is always very close to 1. 
In the other two models (Models 2 and 3), the encounter desorption mechanism is not included and the H$_2$ binding energy is assumed to be either 23~K (Model 2) or 440~K (Model 3).  We note that Model 2 would be very similar to a model where no H$_2$ is allowed to stick on the surface.  Model parameters are summarized in Table~\ref{models}. The chemistry of the disk for the three models is then computed over $5\times 10^6$~yr. For the disk chemical evolution, we keep the same value of the CR ionization rate, although \citet{2013ApJ...772....5C} found that stellar winds can inhibit the propagation of cosmic-ray particles in protoplanetary disks. We explored lower ionization cosmic-rays rates. The quantitative predictions of our models are changed, but not the general conclusions of the paper.

\begin{table}
\caption{Summary of the three models}
\begin{center}
\begin{tabular}{c|c|c}
\hline
\hline
Model & Encounter desorption & H$_2$ binding energy\\
\hline
1 & yes & - \\
2 & no & 23~K \\
3 & no & 440~K\\
\hline
\end{tabular}
\end{center}
\label{models}
\end{table}%

\section{Model results}

\begin{figure}
\includegraphics[width=0.8\linewidth]{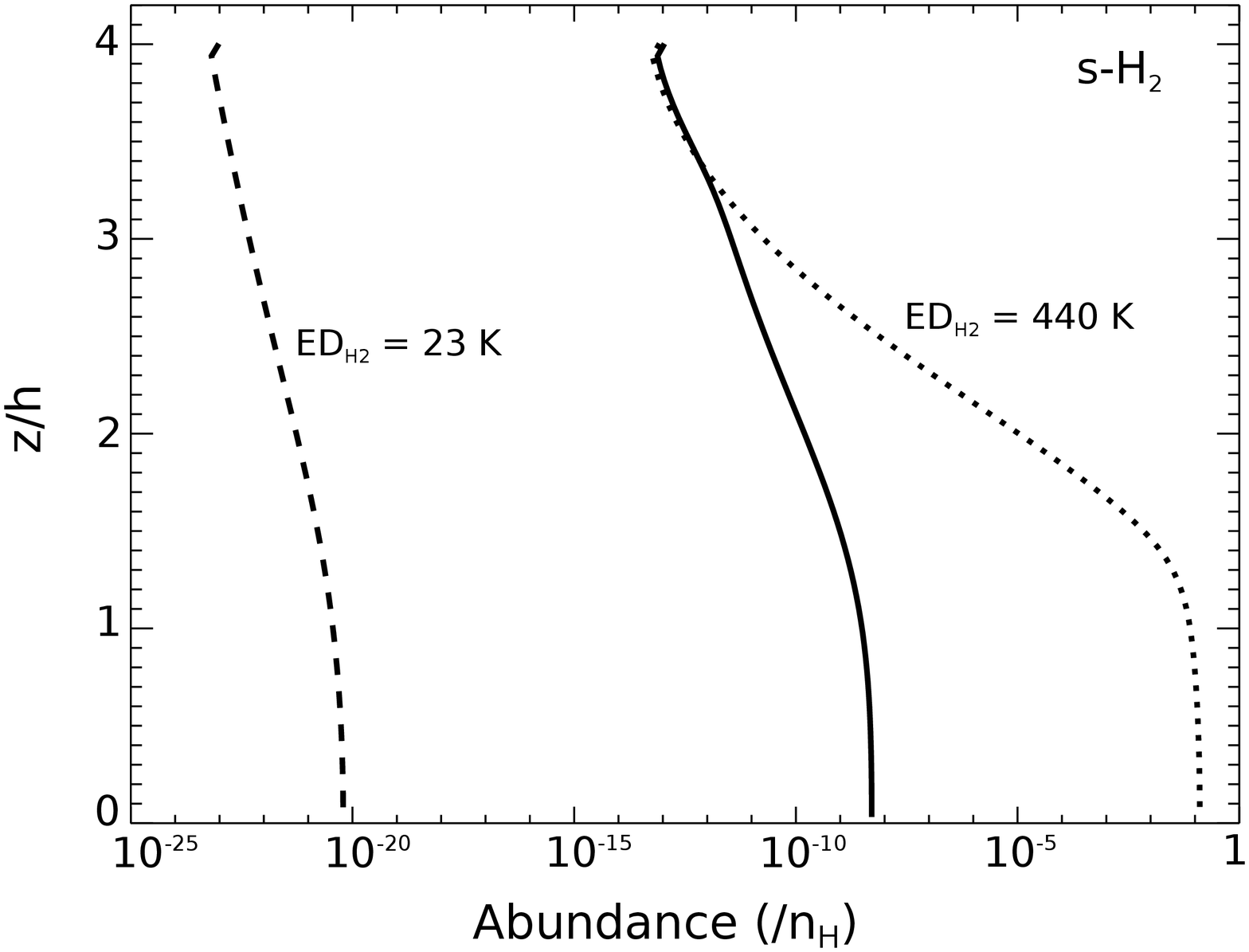}
\includegraphics[width=0.8\linewidth]{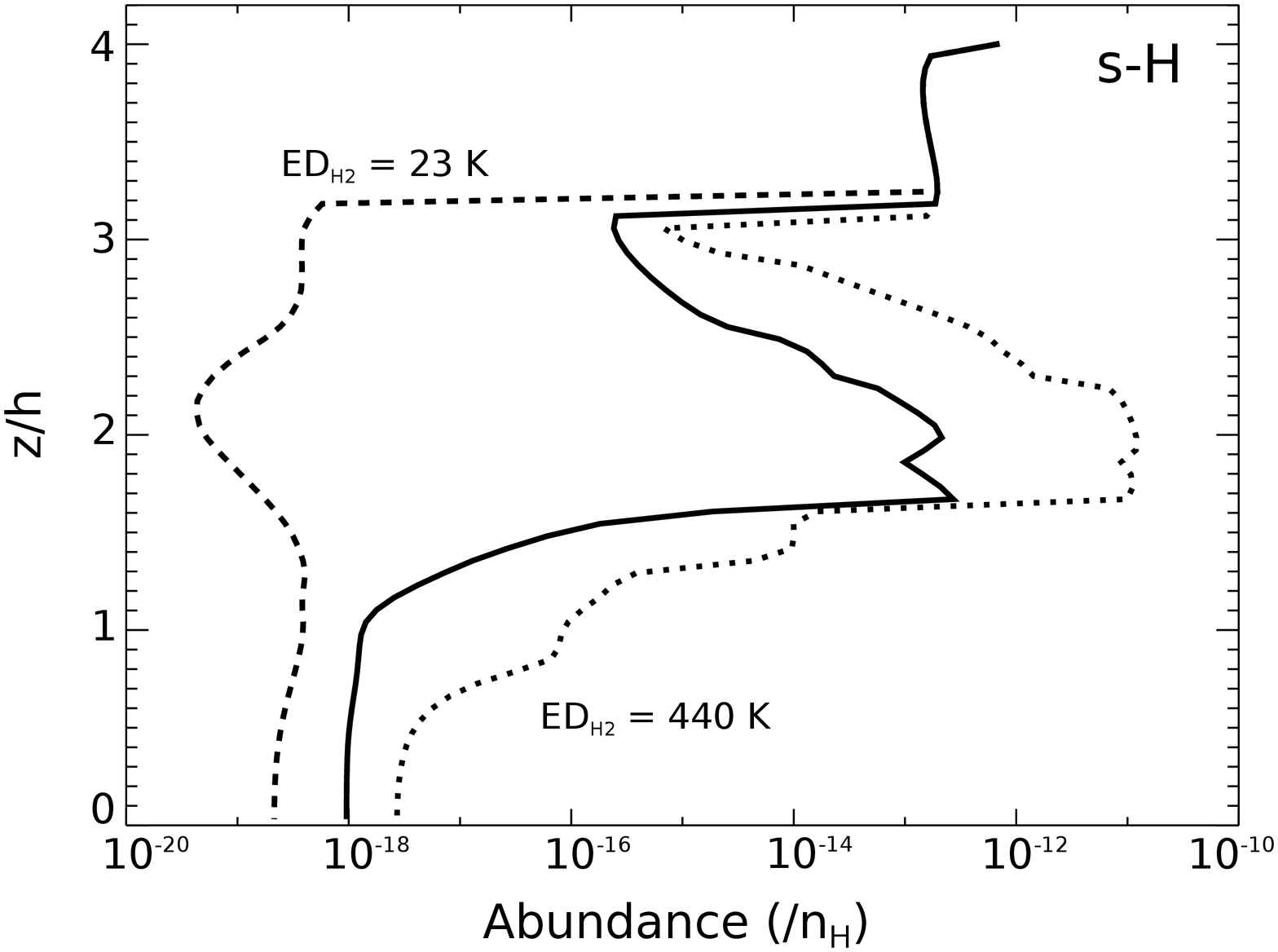}
\caption{Molecular and atomic hydrogen abundance (compared to atomic hydrogen) at grain surfaces as a function of scale height at 300 au. Solid, dashed and dotted lines are for Model 1, 2, and 3, respectively (see Table~\ref{models}). \label{abundances_H}}
\end{figure}

\begin{figure}
\includegraphics[width=0.6\linewidth]{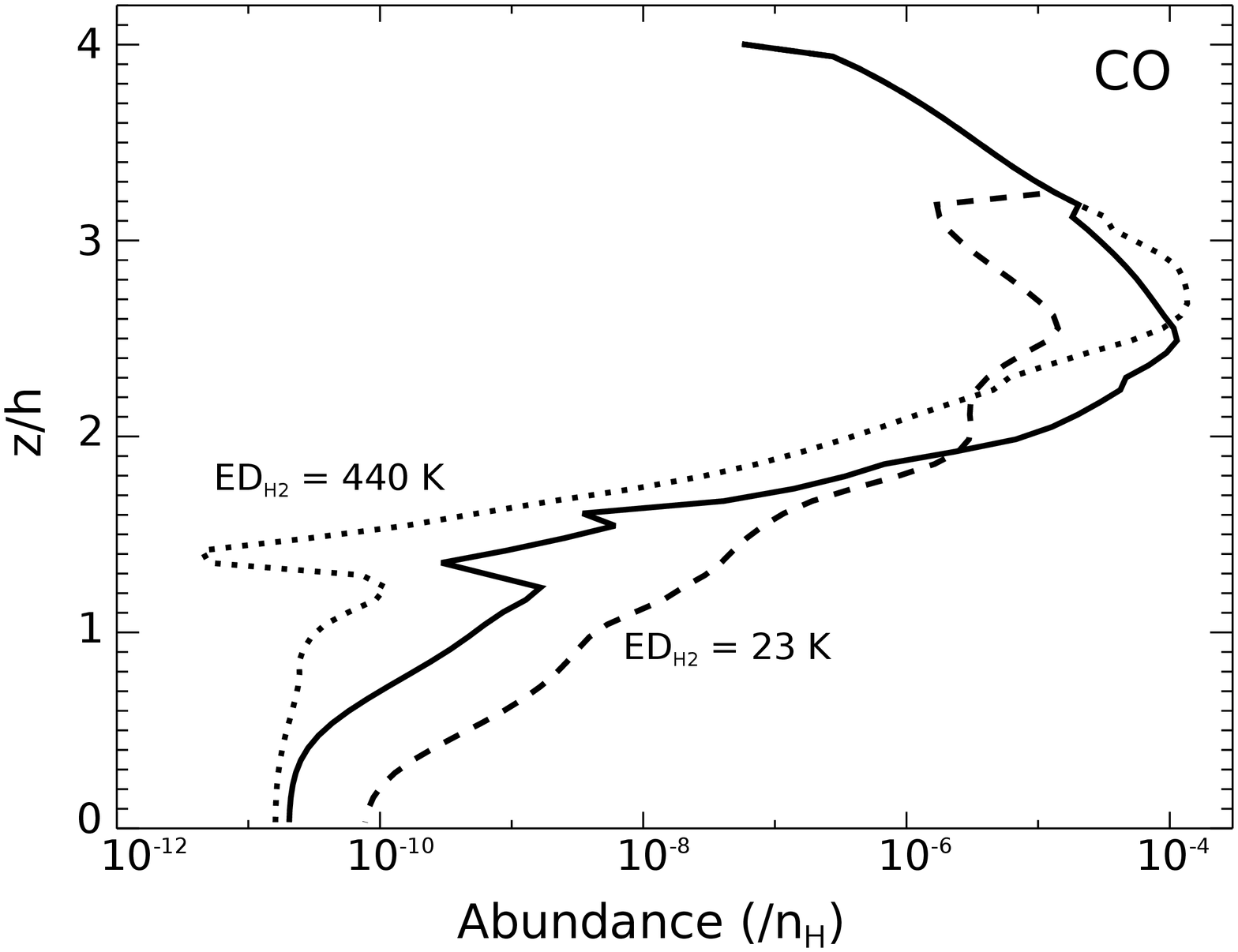}
\includegraphics[width=0.6\linewidth]{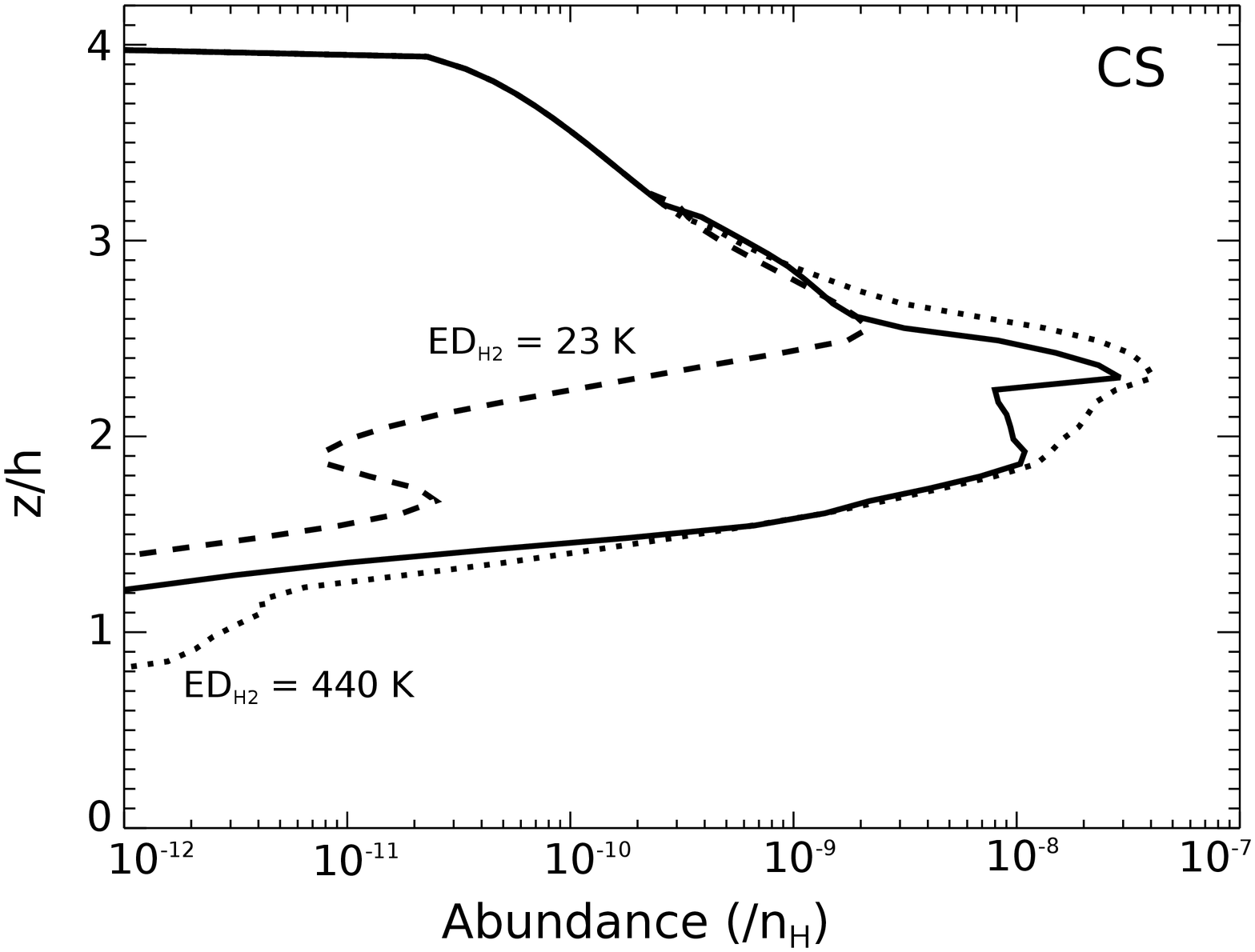}
\includegraphics[width=0.6\linewidth]{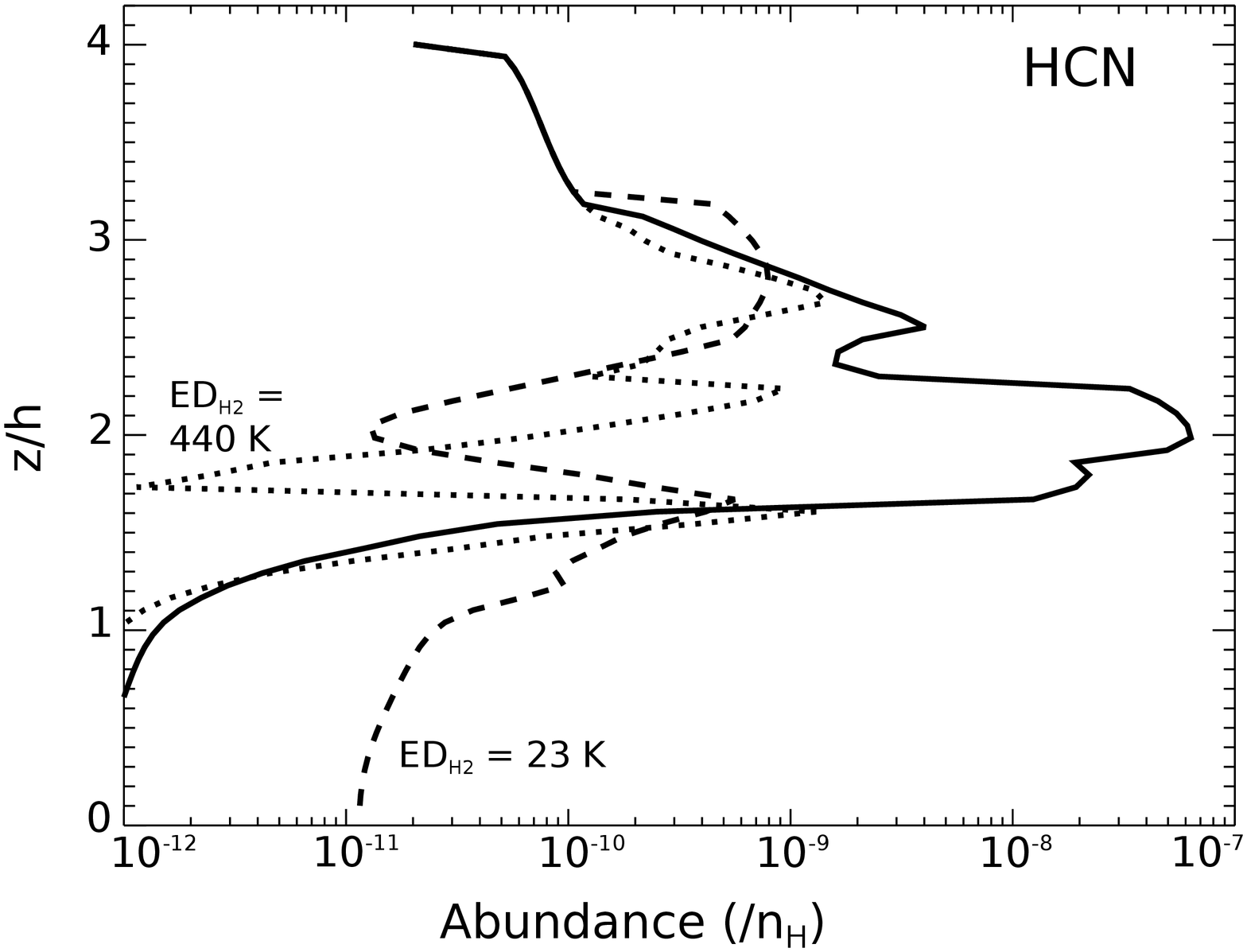}
\includegraphics[width=0.6\linewidth]{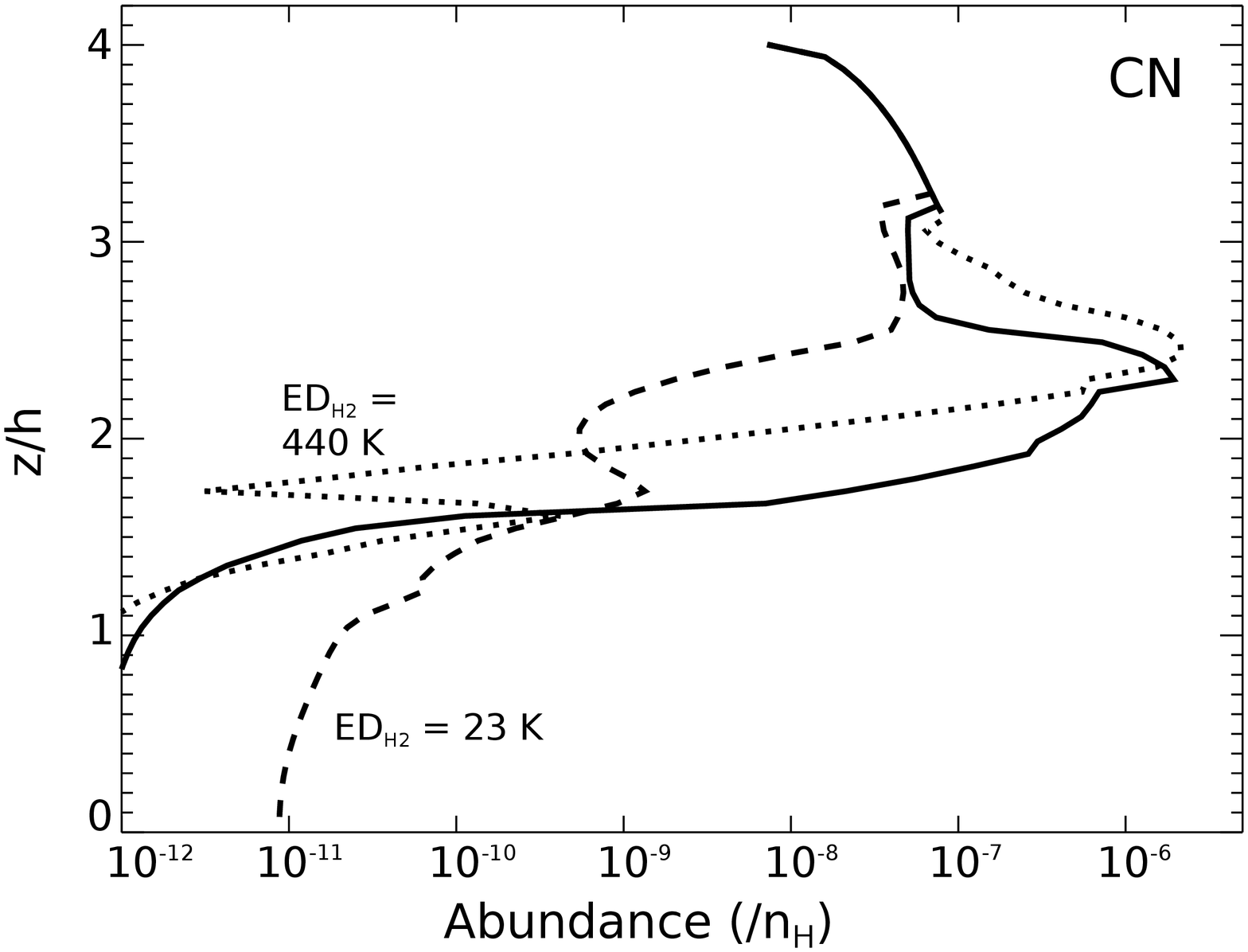}
\caption{Species abundances (compared to H) as a function of scale height at 300 au. Solid, dashed and dotted lines are for Model 1, 2, and 3, respectively (see Table~\ref{models}) \label{abundances_others}}
\end{figure}


\begin{table}
\caption{Species column densities (cm$^{-2}$), in the gas phase at 300 au, integrated in the vertical direction and computed with the three models. Selected species have a Model 1 column density larger than $2\times 10^{12}$~cm$^{-2}$ and a Model 2 or 3 different from Model 1 by more than a factor of 3.}
\begin{center}
\begin{tabular}{l|ccc}
\hline
\hline
Species & Model 1 & Model 2 & Model 3 \\
\hline
Si  & $2\times 10^{12}$  & $2\times 10^{12}$ &  $7\times 10^{11}$\\
Na  & $4\times 10^{12}$ & $4\times 10^{12}$ &  $1\times 10^{11}$\\
OH  & $2\times 10^{14}$ &  $1\times 10^{13}$ & $ 7\times 10^{13}$\\
CO  & $9\times 10^{16}$ &  $7\times 10^{16}$ &  $1\times 10^{16}$\\
CS  & $3\times 10^{13}$ &  $5\times 10^{13}$ &  $1\times 10^{12}$\\
HS  & $5\times 10^{12}$ &  $4\times 10^{12}$ &  $3\times 10^{11}$\\
N$_2$  & $1\times 10^{16}$ &  $8\times 10^{15}$ & $ 1\times 10^{15}$\\
NH &  $5\times 10^{13}$ &  $5\times 10^{12}$ &  $2\times 10^{12}$\\
CH  & $9\times 10^{13}$ &  $7\times 10^{13}$ &   $2\times 10^{13}$\\
O$_2$  & $4\times 10^{12}$ &  $1\times 10^{11}$ & $ 9\times 10^{13}$\\
C$_2$  & $1\times 10^{15}$ &  $4\times 10^{15}$ &  $3\times 10^{13}$\\
CN &  $1\times 10^{15}$ &  $1\times 10^{15}$ &  $3\times 10^{13}$\\
NO &  $2\times 10^{13}$ &  $5\times 10^{12}$ &  $2\times 10^{13}$\\
CCH &  $6\times 10^{13}$ &  $7\times 10^{13}$ &  $7\times 10^{12}$\\
C$_3$ &  $7\times 10^{13}$ &  $1\times 10^{14}$ &  $3\times 10^{12}$\\
CH$_2$  & $4\times 10^{13}$ & $4\times 10^{13}$ &  $1\times 10^{13}$\\
H$_2$O &  $4\times 10^{14}$ &  $2\times 10^{13}$ &  $1\times 10^{13}$\\
HCN &  $1\times 10^{14}$ &  $9\times 10^{12}$ &  $2\times 10^{12}$\\
H$_2$S &  $8\times 10^{12}$ &  $6\times 10^{12}$ &  $8\times 10^{10}$\\
HNC &  $3\times 10^{13}$ &  $3\times 10^{12}$ &  $1\times 10^{12}$\\
HNO  & $3\times 10^{13}$ &  $8\times 10^{12}$ &  $2\times 10^{13}$\\
NH$_2$ &  $5\times 10^{13}$ &  $1\times 10^{13}$  & $2\times 10^{12}$\\
OCN &  $8\times 10^{13}$ &  $6\times 10^{11}$ &  $4\times 10^{12}$\\
C$_4$ &  $2\times 10^{12}$ &  $4\times 10^{12}$ &  $1\times 10^{11}$\\
c-C$_3$H &  $6\times 10^{12}$ &  $9\times 10^{12}$ &  $5\times 10^{11}$\\
l-C$_3$H &  $2\times 10^{12}$ &  $3\times 10^{12}$ &  $4\times 10^{11}$\\
CH$_3$ &  $2\times 10^{14}$ &  $1\times 10^{14}$ &  $4\times 10^{13}$\\
C$_2$H$_2$ &  $5\times 10^{13}$ &  $4\times 10^{13}$ &  $3\times 10^{12}$\\
NH$_3$ &  $5\times 10^{13}$ &  $2\times 10^{13}$ &  $5\times 10^{12}$\\
C$_3$N &  $1\times 10^{13}$ &  $2\times 10^{12}$ &  $1\times 10^{11}$\\
H$_2$CO &  $1\times 10^{13}$ &  $1\times 10^{12}$ &  $2\times 10^{12}$\\
H$_2$CS &  $5\times 10^{12}$ &  $2\times 10^{12}$ &  $1\times 10^{10}$\\
H$_2$CN &  $2\times 10^{12}$ &  $3\times 10^{11}$ &  $1\times 10^{11}$\\
c-C$_3$H$_2$ &  $3\times 10^{12}$ &  $4\times 10^{12}$ & $ 3\times 10^{11}$\\
l
C$_4$H &  $4\times 10^{12}$ &  $6\times 10^{12}$ &  $3\times 10^{11}$\\
CH$_4$ &  $8\times 10^{14}$ &  $6\times 10^{14}$ &  $3\times 10^{13}$\\
HC$_3$N &  $2\times 10^{12}$ &  $8\times 10^{11}$ &  $1\times 10^{10}$\\
H$_2$CCN &  $1\times 10^{13}$ &  $1\times 10^{12}$ &  $7\times 10^{09}$\\
C$_2$H$_3$ &  $5\times 10^{12}$ &  $7\times 10^{12}$ &  $1\times 10^{11}$\\
CH$_2$CCH &  $2\times 10^{12}$ &  $2\times 10^{12}$ &  $2\times 10^{10}$\\
C$_2$H$_4$ &  $1\times 10^{13}$ &  $7\times 10^{12}$ &  $1\times 10^{11}$\\
NC$_4$N &  $2\times 10^{13}$ &  $9\times 10^{12}$ &  $1\times 10^{9}$\\
C$_4$H$_2$ &  $4\times 10^{12}$ &  $6\times 10^{12}$ & $ 2\times 10^{11}$\\
CH$_3$OH  & $4\times 10^{12}$ &  $6\times 10^{11}$ &  $1\times 10^{11}$\\
CH$_2$CHCN &  $2\times 10^{12}$ &  $5\times 10^{10}$ &     $ 2\times 10^{7}$\\
C$_5$H$_2$ &  $4\times 10^{12}$ &  $4\times 10^{12}$ &  $3\times 10^{10}$\\
C$_6$H$_2$ &  $3\times 10^{12}$ &  $3\times 10^{12}$ &  $1\times 10^{10}$\\
C$_2$H$_6$ &  $2\times 10^{15}$ &  $1\times 10^{15}$ &  $5\times 10^{11}$\\
C$_7$H$_2$ &  $3\times 10^{12}$ &  $3\times 10^{12}$ &  $1\times 10^{9}$\\
C$_8$H$_2$ &  $4\times 10^{12}$ &  $4\times 10^{12}$ & $ 9\times 10^{8}$\\
NC$_8$N &  $2\times 10^{12}$ &  $1\times 10^{12}$ &     $ 9\times 10^{5}$\\
C$_9$H$_2$ &  $3\times 10^{12}$ &  $3\times 10^{12}$  &  $   8\times 10^{7}$\\
\hline
\end{tabular}
\end{center}
\label{col_dens}
\end{table}%

Figure~\ref{abundances_H} shows the H$_2$ and H abundances computed at the surface of the grains with the three models. Without the encounter desorption mechanism, the H$_2$ molecule is almost entirely depleted onto the grains (Model 3) or its abundance is very low (Model 2), below z/$h$ = 1. Model 1 is an intermediate case where the abundance of H$_2$ on the grain surfaces is between $10^{-13}$ and $10^{-8}$ depending on the altitude. The s-H abundance (s- stands for physisorbed species at the surface of the grains) presents a similar behavior, but with a peak abundance in the molecular layer (around z/$h$ = 2) in Models 1 and 3. At this altitude, both the atomic and molecular hydrogen abundances differ by several orders of magnitude on the surface depending on the formalism used to treat the sticking of H$_2$ on the grains. In Models 1 and 3, the atomic hydrogen is produced on the surface by the photo-dissociation of water, much less abundant in Model 2 at this altitude, itself formed by the reaction OH + H$_2$ \citep[see][]{2016MNRAS.459.3756R}. \\
The three models also show differences in many other species, even in the gas phase. Figure\ref{abundances_others} shows the vertical abundance profiles for CO, CS, HCN, and CN in the gas phase for the three models. The CO peak abundance in the molecular layer is smaller for Model 2 than for the other two models. In the case of a small abundance of H on the surfaces (Model 2), the physisorbed CO is mostly transformed into CO$_2$. In the other models, CO is hydrogenated on the surfaces and part of the products of the hydrogenation, i.e., HCO, goes back into the gas phase by chemical desorption. HCO is then destroyed by atomic hydrogen and produces CO again, explaining the larger CO gas-phase abundances in Models 1 and 3. The CS molecule is produced in the gas phase by the dissociative recombination of HCS$^+$ in all models. Protonated CS is much less abundant in Model 2 mostly because the carbon is locked in CO and CO$_2$ in the ice. In the other two models, CO in the gas phase is photo-dissociated then C is photo-ionized. C$^+$ participates to the ion-neutral chemistry and to the formation of carbon bearing species. The HCN predicted abundance is very different in Model 1 compared to the two other models at z/$h$ = 2. This molecule is produced in the gas phase mostly by the reaction CN + NH$_3$ $\rightarrow$ NH$_2$ + HCN and NH$_2$ + C $\rightarrow$ HCN + H. All these reactants are much more abundant in Model 1. The CN molecule is formed through different pathways in the three models. In Model 1, it is the photo-dissociation of HCN that is the most efficient. In Model 2, CN is produced on the grains by the reaction C + N, and desorbed by chemical desorption, whereas in Model 3 it is the neutral-neutral reaction N + C$_2$ that is the dominant production reaction. \\
This example illustrates the fact that the gas-phase chemistry in protoplanetary disks, and in particular in the molecular layer where we observe molecules, is strongly affected by the way we treat the surface chemistry, and in particular by the amount of H$_2$ present at the surface of the grains. In Table~\ref{col_dens}, we give the column densities obtained with the three models for abundant species (with a Model 1 column density larger than $10^{12}$~cm$^{-2}$) and that present a difference in column densities computed with Models 2 and 3 larger than a factor of 3 compared to Model 1. The models give completely different predictions for a large fraction of the molecules.  Our model does not include deuterium chemistry but it is very probable that species such as H$_2$D$^+$, which trace the mid-plane, are not affected since H$_3^+$ vertical abundance profiles and surface densities are similar in all three models.  
We also run the models closer to the central star at a radius of 30 au. Here the gas and dust temperatures are high enough to limit the impact of the surface chemistry and the species abundances are then not sensitive to the H$_2$ sticking and abundance on grains.

\section{Conclusions}

In this paper, we modeled the 1D vertical chemistry in the outer part of a DMTau-like proto-planetary disk (at 300 au) using a detailed gas-grain model from \citet{2016MNRAS.459.3756R}. We used three different formalisms to treat the sticking of H$_2$ on the dust grains. In the first model, we used the "encounter desorption" mechanism as described by \citet{2015A&A...574A..24H}, which consists in introducing in the surface network a chemical reaction of the type: s-H$_2$ + s-H$_2$ $\rightarrow$ s-H$_2$ + H$_2$. In other two models, this reaction is not included and the binding energy of H$_2$ is 23~K or 440~K. \\
These three models produce very different H$_2$ and H abundances on the grain mantles, which has a strong impact on the gas-phase chemistry.  In the molecular layer of the disks, the gas-phase molecular abundance of major species, such as CO, CS, HCN or CN, are changed by several orders of magnitude in some cases, as are the vertically integrated column densities. In Model 2, the residing time of H$_2$ on the grain surface is so small that it cannot participate in the surface chemistry. This case is very similar to a model in which H$_2$ would not be allowed to stick at all. In Model 3 on the contrary, the sticking of H$_2$ is so efficient that the molecule disappears from the gas almost completely in the mid-plan. Neither model (2 or 3) is actually physically relevant. From these simulations, it thus appears crucial to treat carefully the sticking of H$_2$ on the grains for protoplanetary disks models. 
The "encounter desorption" mechanism, although perhaps not a fully
physically relevant solution, offers an intermediate step
between models 2 and 3.

\section*{Acknowledgements}

V.W., M.R., F.H, A.D., L.M. and S.G. thank the French CNRS/INSU programme PCMI for their partial support of this work. The research of V.W., M.R., F.H., and L.M. is funded by the ERC Starting Grant (3DICE, grant agreement 336474). V.W. is grateful to Ugo Hincelin and Francois Dulieu for interesting discussions on the encounter desorption mechanism and the determination of H$_2$ binding energies.

%
%

\bibliographystyle{aa}
\bibliography{bib}

\end{document}